\documentclass[apj]{emulateapj}  
\slugcomment{December 2016} 
\usepackage{dcolumn}
\usepackage{bm}
\usepackage{graphicx}
\usepackage{amssymb,amsmath}
\usepackage{epsfig}
\usepackage{color}
\usepackage{wasysym}
\usepackage{natbib}
\usepackage{twoopt}
\usepackage{dashrule}
\definecolor{slateblue}{rgb}{0.1,0.22,0.58}
\usepackage[breaklinks=true,colorlinks=true,linkcolor=slateblue,citecolor=slateblue,urlcolor=slateblue,pdfauthor={Tomassetti},pdftitle={The curious case of CR deuterons}]{hyperref}
\def\MyTitle#1{{\section{#1}}}
\def\Journal#1#2#3#4{{#4}, {#1}, {#2}, #3} 
\definecolor{ColorTitle}{cmyk}{0,.88,.77,.40}

\newcommand{\ApJ}{ApJ}
\newcommand{\AeA}{A\&A}
\newcommand{\PRL}{PRL}
\newcommand{\PRD}{PRD}
\newcommand{\PRC}{PRC}

\newcommand{\ASR}{ASR}

\newcommand{\AMS}{AMS-02}
\newcommand{\etal}{et al.}
\newcommand{\eg}{\textit{e.g.}} 
\newcommand{\ie}{\textit{i.e.}} 

\newcommand{\p}{\textsf{p}}
\newcommand{\pbarp}{\ensuremath{\bar{p}}/\ensuremath{p}}
\renewcommand{\d}{\textsf{d}}
\newcommand{\Hone}{\ensuremath{^{1}}\textsf{H}}
\newcommand{\Htwo}{\ensuremath{^{2}}\textsf{H}}
\newcommand{\Het}{\ensuremath{^{3}}\textsf{He}}
\newcommand{\Hef}{\ensuremath{^{4}}\textsf{He}}

\renewcommand{\H}{\textsf{H}}
\newcommand{\He}{\textsf{He}}

\newcommand{\dHe}{\textsf{d/He}}

\newcommand{\BC}{\textsf{B}/\textsf{C}} 

\newcommand{\Li}{\textsf{Li}} 
\newcommand{\Be}{\textsf{Be}} 
\newcommand{\B}{\textsf{B}} 
\newcommand{\C}{\textsf{C}} 
\newcommand{\N}{\textsf{N}} 
\renewcommand{\O}{\textsf{O}} 
\newcommand{\Oxy}{\textsf{O}} 

\newcommand{\R}{\mathcal{R}}
\renewcommand{\S}{\mathcal{S}}

\usepackage{wasysym}

\usepackage[neverdecrease]{paralist}
\setdefaultleftmargin{0.3cm}{}{}{}{}{}

\begin{document}
\title{The curious case of high-energy deuterons in Galactic cosmic rays}
\author{Nicola Tomassetti$\,^{1}$ and Jie Feng$\,^{2,3}$}
\address{$^{1}$Universit{\`a} degli Studi di Perugia \& INFN-Perugia, I-06100 Perugia, Italy;}
\address{$^{2}$School of Physics, Sun Yat-Sen University, Guangzhou 510275, China;}
\address{$^{3}$Institute of Physics, Academia Sinica, Nankang, Taipei 11529, Taiwan;}

\begin{abstract}
A new analysis of cosmic ray (CR) data collected by the SOKOL experiment in space found that the 
deuteron-to-helium ratio at energies between 500 and 2000 GeV/nucleon takes the value \dHe\,$\sim$\,1.5.
As we will show, this result cannot be explained by standard models of secondary CR production in
the interstellar medium and points to the existence of a high-energy source of CR deuterons. 
To account for the \emph{deuteron excess} in CRs, we argue that the only viable 
solution is hadronic interaction processes of accelerated particles inside old supernova remnants (SNRs).
From this mechanism, however, the \BC{} ratio is also expected to increase at energy above 
$\sim$\,50 of GeV/nucleon, in conflict with new precision data just released by the \AMS{} experiment.
Hence, if this phenomenon is a real physical effect, hadronic production of CR deuterons must occur in 
SNRs characterized by low metal abundance. 
In such a scenario, the sources accelerating \C-\N-\Oxy{} nuclei are not the same as those accelerating 
helium or protons, so that the connection between \dHe{} ratio and \BC{} ratio is broken, and the latter 
cannot be used to place constraints on the production of light isotopes or antiparticles. 
\end{abstract}
\maketitle

\MyTitle{Introduction}  

Deuteron isotopes \Htwo{} are rare particles in Galactic cosmic rays (CRs). 
They are  destroyed rather than formed in thermonuclear reactions in stellar interior
so that, from supernova remnants (SNRs) as sources of CRs, no significant amount of
deuteron is expected to be released by diffusive shock acceleration (DSA).
An important process by which high-energy deuterons can be created in the Galaxy
is nuclear fragmentation of CR nuclei with the gas of the interstellar medium (ISM).
The main source of deuteron production is fragmentation of \Hef{} and \Het{} isotopes, along with
\C-\N-\Oxy{} nuclei, in collisions with interstellar hydrogen and helium.
An important contribution comes from the reaction \p+\p$\rightarrow$\d+$\pi$.
From nuclear fragmentation, interactions of CRs with the ISM are also known to generate
\Li-\Be-\B{} nuclei and antiparticles, that are otherwise rare from stellar nucleosynthesis processes. 
The measured abundances of these elements in the cosmic radiation enables us to  
pose tight constraints on the astrophysical models of CR propagation in the Galaxy and, 
in the context of indirect searches of dark matter, to asses the astrophysical background of CRs antiparticles \citep{Grenier2015}.

Very recently, the Alpha Magnetic Spectrometer (\AMS) experiment has reported a new precise measurement of 
the \BC{} ratio in CRs at energy between $\sim$\,0.5 and 1000\,GeV/n \citep{Aguilar2016BC}.
At kinetic energies above  $\sim$\,10\,GeV/n, the ratio is found to decrease steadily with increasing energy. 
Since boron nuclei are mainly produced by \C-\N-\O{} fragmentation in the ISM,
the observed trend of the \BC{} ratio reflects the conception that CRs diffusively propagate
in the Galactic magnetic fields with an average diffusion coefficient that increases with energy.
According to this picture, the ratio between \Htwo{} and its main progenitor \Hef{} 
must follow the same behavior in the GeV-TeV energy range. 
Recent calculations have shown, indeed, that the observed boron and deuteron abundances at GeV/n 
energies can both be self-consistently described by diffusion models of CR propagation \citep{Coste2012,Tomassetti2012Iso}.
In the GeV-TeV energy region, the \dHe{} ratio is expected to decrease rapidly, as fast as the \BC{} ratio does,
but CR deuterons have never been detected at these energies.

Quite unexpectedly, a new analysis of the data collected by the satellite mission SOKOL
has determined the \dHe{} ratio at 0.5-2 TeV/n energy \citep{Turundaevskiy2016}. 
In this measurement, the discrimination between $^{1,2}$\H{} and $^{3,4}$\He{}
isotopes was performed by means of neural network analyses of the topology of hadronic showers developed by these particles. 
By means of two different Monte Carlo simulations, consistent results have been obtained:
0.114$\pm$0.023 and 0.099$\pm$0.021 for the \Htwo/(\Hone$+$\Htwo) ratio,
1.64$\pm$0.30 and 1.43$\pm$0.27 for the \Htwo/\Hef{} ratio. 

In this Letter, we show that the above results represent a striking anomaly in CR physics that cannot be explained 
by standard models of CR propagation and secondary production in the ISM. 
Then, we argue that a deuteron excess can be explained in terms of
hadronic production occurring inside SNRs, which was proposed in \citet{Blasi2009}
to explain the positron excess in CRs \citep[see also][]{Kachelriess2011,Serpico2012}.
Using new  evaluations of fragmentation cross-sections, we calculate 
for the first time the high-energy production of CR deuterons in SNRs and in the ISM, demonstrating that
this mechanism can account for the new \dHe{} data. 
Along with the \dHe{} ratio, we also compute the \BC{} ratio under the same framework,
showing that there are conflicting results in  the model predictions for the two observables.
We therefore conclude that this tension can be resolved if the deuterons
progenitors are not accelerated in the same sources of boron progenitors. 
We discuss our results and their implications
for the interpretation of antiproton data.
\\

\MyTitle{Calculations}    
\label{Sec::Calculations} 

We compute the spectrum of CR nuclei accelerated in SNRs within the linear DSA theory and including the production of secondary fragments. 
Similar calculations are done in earlier works \citep{Blasi2009,BlasiSerpico2009,MertschSarkar2009,MertschSarkar2014,TomassettiDonato2015,Herms2016}.  
We follow closely the derivation of \citet{TomassettiDonato2012}.
In the shock rest-frame  ($x = 0$), the upstream plasma flows in from $x < 0$ with speed $u_{1}$ (density $n_{1}$) and the downstream plasma flows 
out to $x > 0$ with speed $u_{2}$ (density $n_{2}$). The compression ratio is $r=u_{1}/u_{2} = n_{2}/n_{1}$. 
For a nucleus with charge $Z$ and mass number $A$, the equation describing diffusion and convection at the shock reads
\begin{equation} \label{Eq::DiffusionDSA}
 u \frac{\partial f}{\partial x} = D \frac{\partial^{2}
 f}{\partial x^{2}} + 
 \frac{1}{3}\frac{du}{dx}p\frac{\partial f}{\partial p} 
 -\Gamma^{\rm tot}{f}  + Q \,,
\end{equation}
where $f$ is the phase space density,  $D(p)$ is the diffusion coefficient near the shock, $u$ is the fluid speed,
and $\Gamma^{\rm tot} = c n \sigma^{\rm tot}$ is the total fragmentation rate for
cross-sections $\sigma^{\rm tot}$ and background density $n$, which is assumed to be composed of \H{} and \He{} like the average ISM. 
The source term includes particle injection at the shock, $Q = Y \delta(x) \delta(p-p^{\rm inj})$,
with $p^{\rm inj}\equiv Z R^{\rm inj}$ and $R^{\rm inj}\cong 0.5\,$GV for all nuclei.
The $Y$-constants set the normalization of each species.
We assume strong shocks ($r\approx 4$) and a diffusion coefficient $D = \kappa_{B}\frac{p/Z}{3B}$, where $\kappa_{B}$ 
parameterizes the deviation of $D(p)$ from the Bohm value due to magnetic damping.
The resulting acceleration rate at momentum $p$  is $\Gamma^{\rm acc} \sim u_{1}^{2}/20\,D$.
For an SNR of age $\tau_{\rm snr}$, the condition $\Gamma^{\rm acc}=\tau^{-1}_{\rm snr}$ defines the maximum momentum scale attainable by DSA. 
In the presence of hadronic interactions, the additional requirement $\Gamma^{\rm tot} \ll \Gamma^{\rm acc}$ must be fulfilled. 
The downstream solution reads 
\begin{equation}\label{Eq::DownStreamSecondary}
f_{2}(x,p) = f_{0}(p) \left( 1 -  \frac{\Gamma^{\rm tot}_{2}}{u_{2}}x \right) + \frac{q_{2}}{u_{2}}x \,,
\end{equation}
where $f_{0}(p)$ is the distribution function at the shock. As found in \citet{MertschSarkar2009}, $f_{0}(p)$ is given by
\begin{equation}\label{Eq::FullSolutionAtShockFront} 
  \begin{aligned} 
f_{0}(p) =  \alpha \int_{0}^{p} \left( \frac{p'}{p} \right)^{\alpha} Y\delta(p'-p^{\rm inj})  e^{-\chi(p,p')} \frac{dp'}{p'}\\
+ \alpha \int_{0}^{p} \left( \frac{p'}{p} \right)^{\alpha} \frac{q_{1} D}{u^{2}_{1}}\left( 1 + r^{2} \right) e^{-\chi(p,p')} \frac{dp'}{p'} \,,
\end{aligned}
\end{equation}
with $\alpha = 3r/(r-1)$,
$\chi \approx  \alpha(\Gamma^{\rm tot}_{1}/\Gamma^{\rm acc}) [ D(p) - D(p') ]$, and
the subscript $i=1$ ($i=2$) indicates the upstream (downstream) region. 
The first term of Eq.\,\ref{Eq::FullSolutionAtShockFront} describes primary particles injected 
at the shock and it is of the form $\sim p^{-\alpha} e^{-\chi}$. 
The second term of Eq.\,\ref{Eq::FullSolutionAtShockFront} describes the production and acceleration of CR fragments 
from heavier progenitors. For each $k\rightarrow j$ process, the $q$--term of 
Eq.\,\ref{Eq::FullSolutionAtShockFront} is given by $q_{kj}^{\rm sec}(p) = \xi_{kj}^{-3} f_{j}(x,p/\xi_{kj}) \Gamma^{\rm fr}_{kj}$,
where $\Gamma^{\rm fr}_{kj}= \sigma_{kj}n \beta c$ is the secondary production rate, and $\sigma_{kj}$ is the corresponding cross-section. 
The momentum inelasticity factor $\xi_{kj} = A_{j}/A_{k}$ expresses the conservation of kinetic energy per nucleon between progenitor and fragment. 

\begin{figure}[!t]
\epsscale{1.09}
\plotone{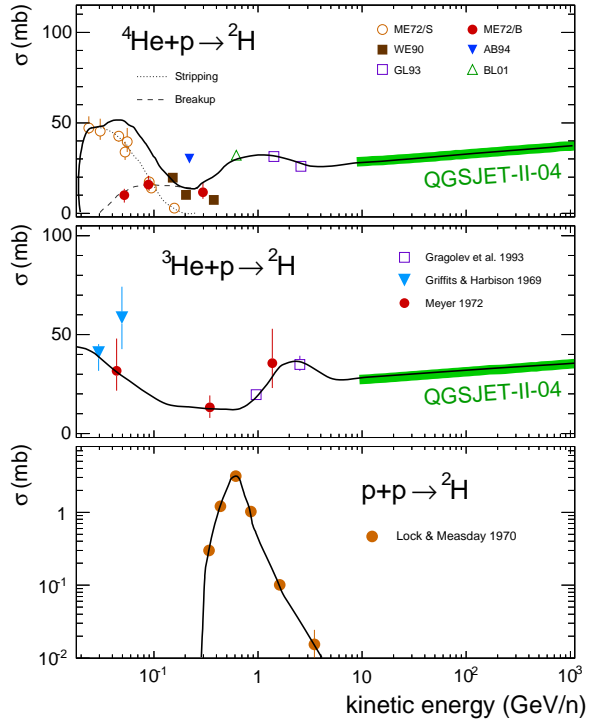}
\caption{ 
Cross-section data for deuteron production from \Hef-\p, \Het-\p, and \p-\p{}
collisions along with new calculations obtained from QGSJET-II-04 and our parameterizations. 
Data are from the compilations in \citet{Tomassetti2012Iso} and \citet{Coste2012}.
}
\label{Fig::ccHeliumDeuteronXS}
\end{figure}

Equation\,\ref{Eq::DiffusionDSA} is solved for all relevant species.
We considered primary nuclei (with $Y>0$) \p, \Hef, \C, \N, \O{} tuned to recent data as in \citet{MertschSarkar2014}, 
and secondary \B{} production from \C-\N-\O{} collisions with hydrogen and helium gas.
The adopted fragmentation cross-sections are those re-evaluated in \citet{Tomassetti2015XS}.
We account for deuteron production from collisions of \p, \Het, and \Hef{} off hydrogen and helium.
Measurements and calculations for these reactions are available only below 10\,GeV/n of energy. 
Thus, we have performed new calculations at $E=$10-1000\,GeV/n using the hadronic Monte Carlo 
generator \texttt{QGSJET-II-04} \citep{Ostapchenko2011}.
The cross-sections for the dominant channels are shown in Fig.\,\ref{Fig::ccHeliumDeuteronXS}.
The solid lines are our parameterizations adapted from \citet{Tomassetti2012Iso} and extended to higher energy.
Cross-sections for collisions off \He{} target are larger by nearly a factor of two and contribute to $\sim$20\,\%.
The contribution from fragmentation of heavier nuclei is negligible. 
In all reactions, kinetic energy per nucleon is approximately conserved
except for the \p$+$\p{} reaction, which peaks at a proton energy of $\sim$\,600\,MeV/n
and produces deuterons within a narrow energy range around 150\,MeV/n \citep{Meyer1972}.
Hence, these deuterons experience DSA acceleration similarly to 
primary components with non-zero $Y$-factor, \ie, to a power-law spectrum $\sim p^{-\alpha}$.
%
In contrast, deuterons ejected by helium fragmentation have an initial spectrum $p^{-\alpha}$ 
so that, for those produced within a distance $d\lesssim\,D/u_{1}$ from the shock and undergo DSA,
their final spectrum reaches the form 
$f^{\rm sec}_{0}\sim f^{\rm pri}_{0}D(p) \propto p^{-\alpha+1}$ (see Eq.\,\ref{Eq::FullSolutionAtShockFront}).
This means that, for Bohm-type diffusion, the secondary deuteron spectrum is \emph{one power harder} than that 
of its progenitors. The total CR flux produced by SNRs is computed as
\begin{equation}\label{Eq::SNRVolumeIntegral}
\S^{\rm snr}(p) = 4\pi p^{2}\mathcal{R_{\rm SN}} \int_{0}^{\tau_{\rm snr}u_{2}} 4 \pi x^{2} f_{2}(x,p) dx
\end{equation}
where $\R_{\rm SN}\cong$\,25\,Myr$^{-1}$\,kpc$^{-2}$ is the explosion rate per unit volume
and $\tau_{\rm snr}\cong$\,40\,kyr is the age of the SNR.

To model the subsequent propagation of CRs in the ISM, 
we adopt a two-halo model of CR diffusion and nuclear interactions \citep{Tomassetti2012,Tomassetti2015TwoHalo}.
The Galaxy is modeled as a disk of half-thickness $h\cong$\,100\,pc 
containing SNRs and gas with number density $\tilde{n}\cong$\,1\,cm$^{-3}$. 
The disk is surrounded by a diffusive halo of half-thickness $L$ and zero matter density. 
We give a one-dimensional description in the thin disk limit ($h \ll L$). 
For each CR nucleus, the transport equation reads
\begin{equation}\label{Eq::Transport1D}
\frac{\partial \mathcal{N}}{\partial t} = \frac{ \partial}{\partial z} \left( K(z) \frac{\partial \mathcal{N}}{\partial z} \right) 
-2h \hat{\delta}(z) \tilde{\Gamma}^{\rm tot} \mathcal{N} + 2h\hat{\delta}(z) \S^{\rm tot} \,,
\end{equation}
where $\mathcal{N}$ is its density,
$\hat{\delta}$ is a Dirac function of the $z$-coordinate, $K(z)$ is the diffusion coefficient
of CRs in the Galaxy, and $\tilde{\Gamma}^{\rm tot} = \beta c \tilde{n} \sigma^{\rm tot}$ 
is the destruction rate in the ISM at velocity $\beta c$ and cross-section $\sigma^{\rm tot}$. 
The source term $\S^{\rm tot}$ is split into a primary term $\S^{\rm snr}$, obtained from Eq.\,\ref{Eq::SNRVolumeIntegral}
as solution of the DSA equation, and a secondary production term $\S^{\rm sec}= \sum_{k} \tilde{\Gamma}_{k}^{\rm fr} \mathcal{N}_{k}$,
from fragmentation of $k$-type nuclei in the ISM with rate $\tilde{\Gamma}_{k}^{\rm fr}$.
To compute the interaction rates in the ISM $\tilde{\Gamma}^{\rm in/fr}$ we adopt
the same fragmentation network (and same cross-sections) as occurring inside SNRs.
Equation\,\ref{Eq::Transport1D} is solved in steady-state conditions $\partial \mathcal{N}/\partial t =0$.
The derivation of the full solution is in \citet{Tomassetti2012}.
The diffusion coefficient is taken of the type $K(p,z) = \beta K_{0}((pc/Ze)/GV)^{\delta(z)}$ where 
$K_{0}$ expresses its normalization.
For the scaling index, $\delta(z)$, we adopt $\delta=\delta_{0}$ in the region of $|z|<\xi\,L$ (inner halo)
and  $\delta=\delta_{0}+\Delta$ for $|z|>\xi\,L$ (outer halo). 
Our default parameters are set as  $\xi\,L\cong\,0.1$\, $\delta_{0}\cong$\,1/3, 
$\Delta\cong$\,0.55, and $K_{0}/L\cong$0.01\,kpc\,Myr$^{-1}$. 
The differential energy fluxes of each species are given by $J(E) = \frac{\beta c}{4 \pi}\mathcal{N}$.
Solar modulation is described in \textit{force-field} approximation 
using the parameter $\phi=500$\,MV for a medium-level modulation strength. 
The \dHe{} and \BC{} ratios as function of kinetic energy per nucleon are eventually calculated 
as $J_{^{2}{\rm H}}/J_{^{4}{\rm He}}$ and $(J_{^{10}{\rm B}}+J_{^{11}{\rm B}})/(J_{^{12}{\rm C}}+J_{^{13}{\rm C}})$, respectively.
In the following, we consider two model implementations representing two alternative scenarios:
\\[0.05cm]
\begin{itemize}
\item \emph{Scenario \#\,1 (B/C-driven, conservative)} --- standard model without interactions in sources, 
  which is the case of a CR flux released by \emph{young} SNRs with amplified magnetic fields
  ($B\gtrsim$\,100\,$\mu\,G$) and/or low background density ($n_{1}\approx\,10^{-3}$\,cm$^{-3}$).
  In this model, secondary production of CRs deuterons or \Li-\Be-\B{} occurs only in the ISM. 
  This model is tuned to match the new \BC{} data from \AMS{} at GeV/n-TeV/n energies.
\item \emph{Scenario \#\,2 (d/He-driven, speculative)} --- model with copious production and acceleration of secondary particles in SNR shockwaves,
  which is the case for a GeV-TeV flux provided by \emph{old} SNRs with damped magnetic fields, slow shock speed, or dense
  ambient medium, \ie, with the combination $n_{1}\kappa_{B}B^{-1}u_{8}^{-2}\sim$\,400.
  This model is tuned against the \dHe{} data including the new SOKOL measurement at TeV/n energies.
\end{itemize}

\begin{figure*}[!t]
\epsscale{0.48}
\plotone{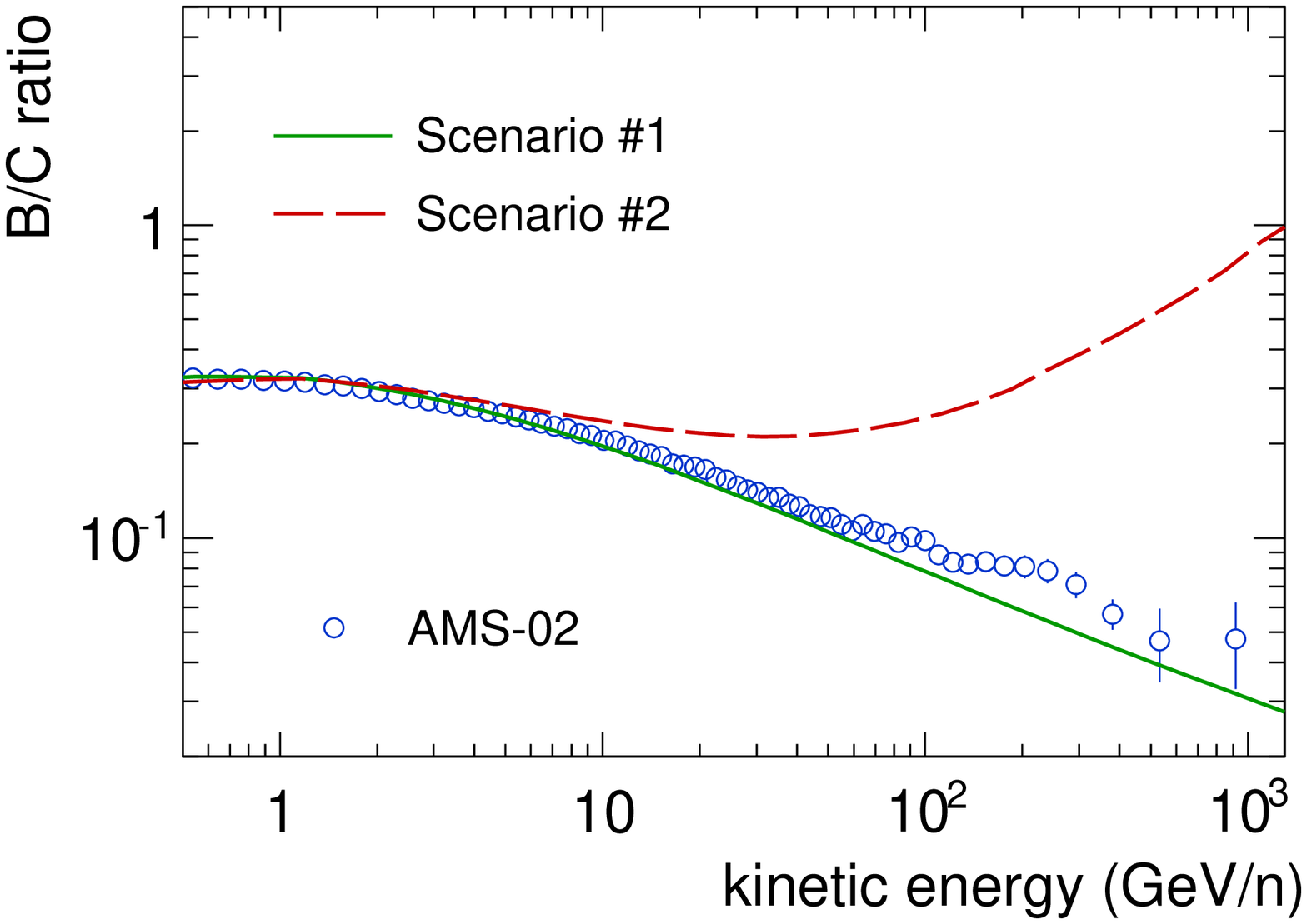}
\qquad
\qquad
\qquad
\plotone{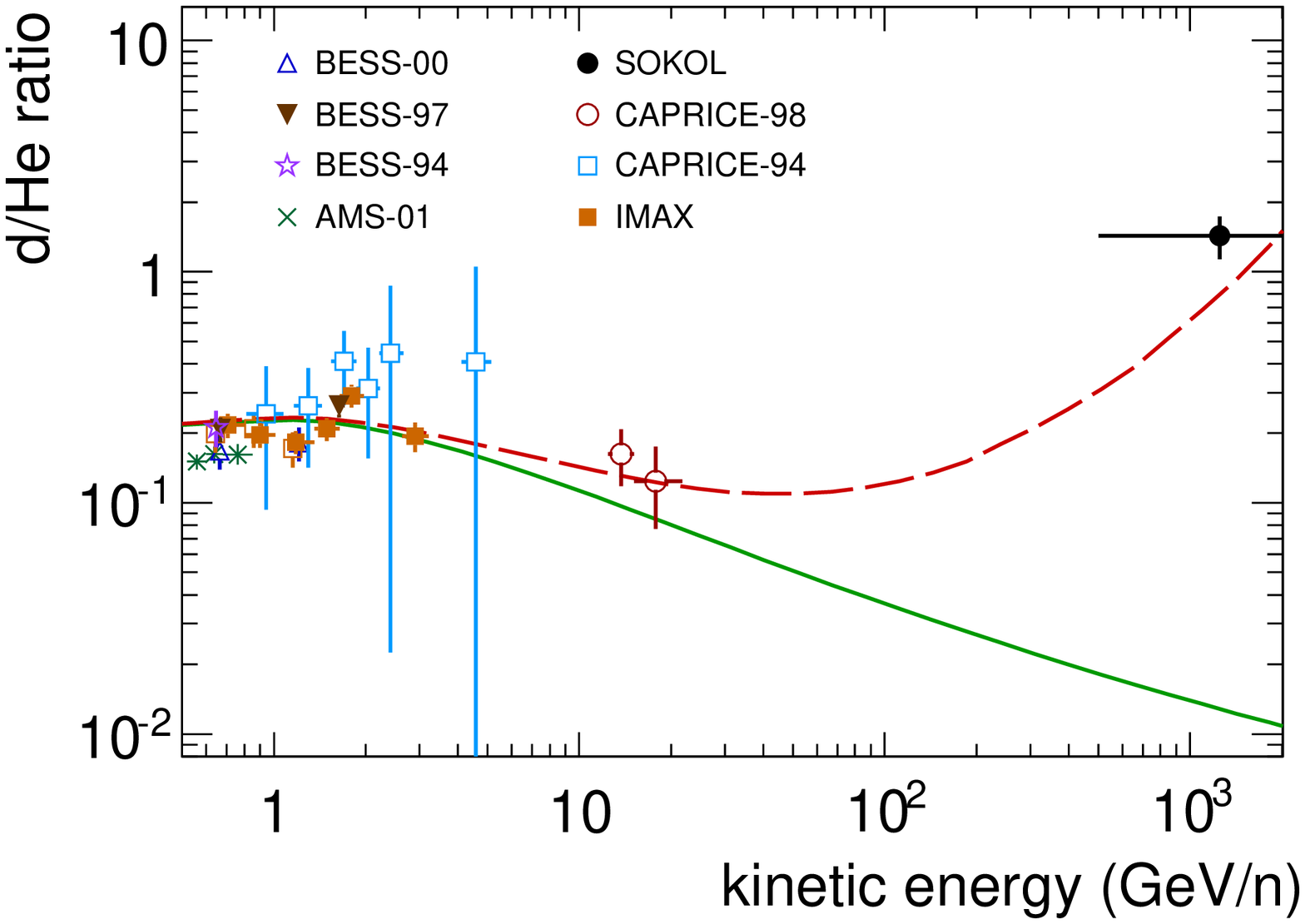}
\caption{
  Model calculations from \emph{Scenario\,\#\,1} (green solid lines) and
  \emph{Scenario\,\#\,2} (red dashed lines) for the \BC{} ratio (left) and \dHe{} ratio (right).
  Data are from \AMS{} \citep{Aguilar2016BC}, BESS \citep{Myers2005,Kim2013}, CAPRICE \citep{Papini2004}, 
  IMAX \citep{DeNolfo2000}, AMS-01 \citep{Aguilar2011Iso}, and SOKOL \citep{Turundaevskiy2016}.
}\label{Fig::ccDtoHeRatio}
\end{figure*}

\MyTitle{Results and Discussion}  

Model calculations are shown in Fig.\,\ref{Fig::ccDtoHeRatio} for the \BC{} ratio the \dHe{} ratio 
at energies between $\sim$\,0.5\,GeV and 2\,TeV per nucleon.
\emph{Scenario\,\#\,1} is plotted as green solid lines.
In this model, secondary nuclei such as \d{} or \B{} are entirely generated in the ISM, \ie, without SNR components,
and thus secondary/primary ratios decrease steadily as 
$J_{s}/J_{p}\propto (L/K_{0})\left[ \xi + (1-\xi)\rho^{-\Delta} \right]\rho^{-\delta_{0}}$
where $\rho=(pc/Ze)/GV$.
In the high-energy limit one has $J_{s}/J_{p} \propto E^{-1/3}$. 
It can be seen that this model fits remarkably well the new \AMS{} data on the \BC{} ratio 
and dictates a similar trend for the \dHe{} ratio, 
which is predicted to reach the level of $\sim\,10^{-2}$ in the TeV/n energy scale.
We therefore conclude that the SOKOL measurement of the \dHe{} ratio 
is \emph{at least two orders of magnitude} higher than that expected from
standard models where CR deuterons are produced by fragmentation in the ISM.

In \emph{Scenario\,\#\,2}, shown as red dashed lines, 
hadronic interaction processes inside SNRs generate a source component of secondary nuclei,
which is harder than that arising from CR collisions with the ISM and, as discussed,
even harder than that of primary \p-\He{} spectra.
It is then possible, with fragmentation inside SNRs, having secondary/primary ratios
that \emph{increase}  with energy. 
Figure\,\ref{Fig::ccDtoHeRatio} shows that \emph{Scenario\,\#\,2} matches fairly well 
the \dHe{} ratio measurements at
GeV/n and at TeV/n energies, therefore providing an explanation for the new SOKOL data.
Under this model, however, the \BC{} ratio is also predicted to increase, at energies 
above $\sim$\,50\,GeV/n, in remarkable contrast with the new \AMS{} data.
While interactions inside SNRs seem to be the only mechanism capable of explaining a rise in the
\dHe{} ratio, it is apparent that the observed decrease of the \BC{} ratio conflicts with this mechanism.
We also note that, at the $\sim$\,1\,GeV/n energy region where secondary CR production in the ISM dominates,
the two ratios are consistently described by both models \#\,1 and \#\,2,
at least within the precision of the current data.

As we see it, the only solution to this tension is a situation where \emph{the connection between \dHe{} ratio and \BC{} ratio is broken}.
This situation is realized if the sources accelerating helium and protons (and producing deuterons) 
are not the same as those accelerating heavier \C-\N-\O{} nuclei and,
in particular, deuterons must be accelerated by a low-metallicity source,
which may be the case for SNRs expanding over \H-dominated molecular clouds. 
Such a possibility was also discussed in \citet{CholisHooper2014} (see Sect.\,VI),
and proposed in other works \citep{Fujita2009,Kohri2016}, all focused on antiparticle excesses in CRs.
In such a scenario, the \BC{} ratio can no longer be used to place constraints on antimatter spectra.
In this respect, it is important to note that the connection between \dHe{} ratio 
and antimatter/matter ratios would still be preserved because, 
in contrast to \Li-\Be-\B{} nuclei, secondary deuterons share their progenitors with positrons and antiprotons.
%
\begin{figure}[!t]
\epsscale{1.0}
\plotone{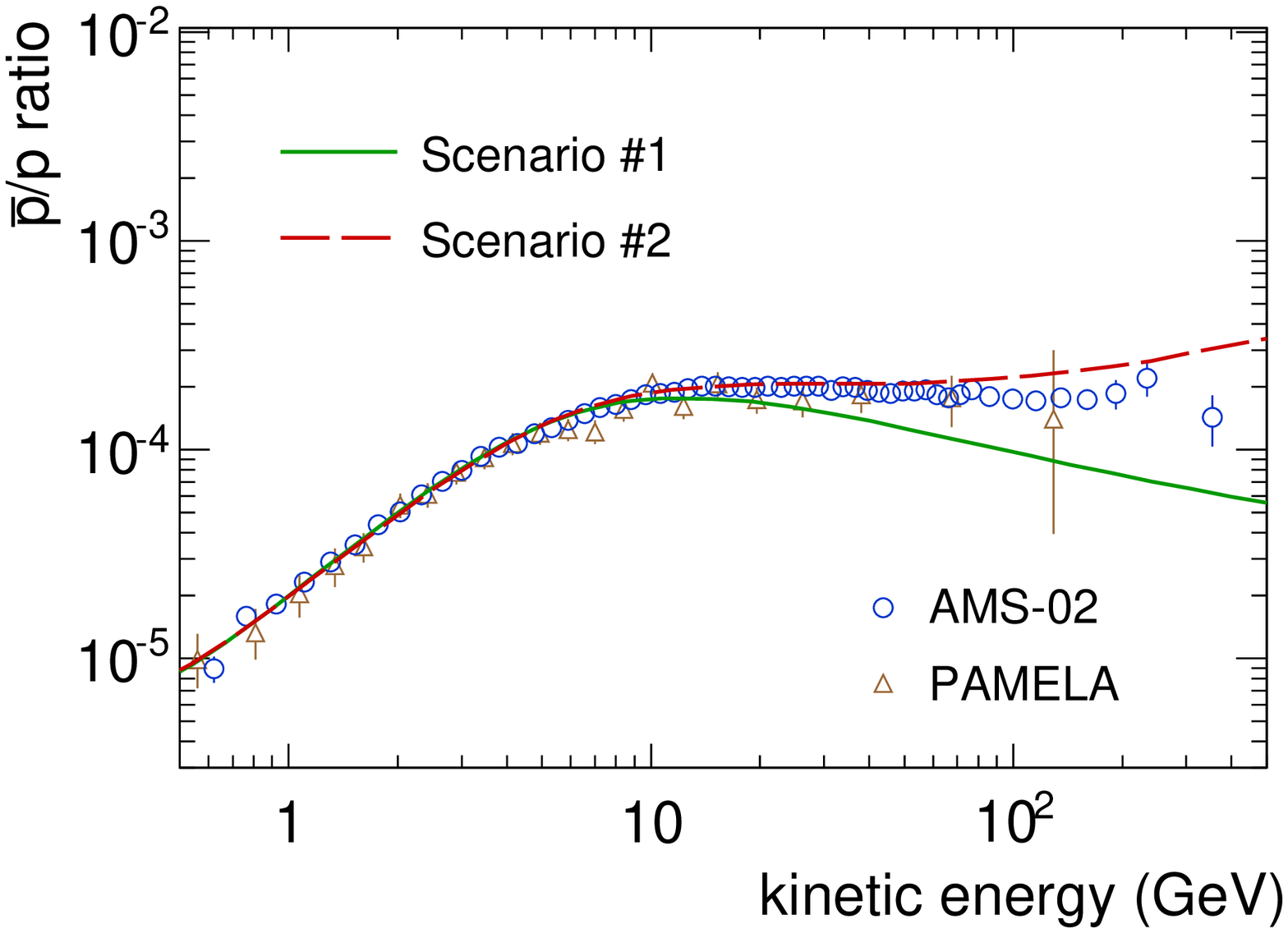}
\caption{ 
Antiproton/proton ratio from our calculations in comparison with new data from \AMS{} and PAMELA\,\citep{Adriani2010PbarP,Aguilar2016PbarP}.
}
\label{Fig::ccPbarPRatio}
\end{figure}
%
It is then interesting to calculate the antiproton/proton ratio 
in light of new data released by \AMS{} \citep{Aguilar2016PbarP}.
Calculations were performed in \citet{BlasiSerpico2009} and subsequent works \citep[\eg,][]{MertschSarkar2014}. 
In this work, the antiproton distribution at the shock $f_{0}(p)$ is calculated numerically, as done in
\citet{Herms2016}, in order to drop the ``inelasticity approximation''
that links the antiproton momentum $p$ to the primary proton momentum $p_{p}$ through an assumedly constant factor $\xi\equiv\langle p/p_{p}\rangle$.
It was noted that such an approximation leads to an overestimate of the high-energy antiproton production in SNRs \citep{Kachelriess2011}.
Along with antiproton production from \p-\He{} collisions with hydrogen and helium gas, 
we also account for \emph{tertiary} reactions (such as $\bar{p}+p\rightarrow\bar{p}^{\prime}+X$) and for destruction processes. 
All these processes are implemented in both acceleration and propagation. The corresponding cross-sections
are taken from \citet{Jie2016}.
The resulting \pbarp{} ratio is shown in Fig.\,\ref{Fig::ccPbarPRatio}.
It can be seen that Scenario\,\#2 is preferred by the \AMS{} data.
Reducing nuclear uncertainties in antiproton production is clearly essential for
a complete discrimination between the different models \citep{Jie2016}.
\\
 
\MyTitle{Experimental Challenges}  

Given the implications of these new data on the phenomenology of CR propagation, 
we believe that the situation deserves more clarification on the experimental side.
The SOKOL analysis relies on unconventional techniques of deuteron/proton mass separation,
which is always a very challenging task. 
For instance the \dHe{} measurement might be overestimated due to undetected background arising, \eg, from the
mass distribution tails of CR protons or from \Hef{} nuclei fragmenting in the top of the instrument.
On the other hand, it is very unlikely for such a background to affect the results by two orders of magnitude.
The deuteron spectrum and the \dHe{} ratio are being precisely measured by the \AMS{} 
experiment at $E\sim$\,0.1-10 GeV/n with standard spectrometric techniques.
In addition, \AMS{} is also equipped with a Transition Radiation Detector (TRD), designed for lepton/hadron mass separation, 
which can provide direct measurements of the Lorentz factor $\gamma$ at TeV/n energies \citep{Obermeier2015}.
In standard magnetic spectrometers, the CR mass is derived from velocity and momentum measurements,
$M\propto\,p/(\gamma\beta c) = \frac{p}{\beta c}\sqrt{1-\beta^{2}}$, so that its corresponding resolution $\delta M/M$ is given by
\begin{equation}\label{Eq::MassResolution}
\left(\frac{ \delta M}{M}\right)^{2} = \left( \frac{\delta p}{p} \right)^{2} + \gamma^{4}\left( \frac{\delta\beta}{\beta} \right)^{2},
\end{equation}
showing that the mass resolution degradates rapidly, at relativistic energies, due to the $\gamma^{4}$-factor.
In contrast, with 
the opportunity of performing
direct TRD-based $\gamma$-measurements, \AMS{} may have the capability
to detect CR deuterons at the $\mathcal{O}({\rm TeV})$ energy scale.
\\

\MyTitle{Conclusions}  

This work is aimed at interpreting new data, registered by the SOKOL experiment in space,
that revealed a surprisingly high abundance of CR deuterons in the TeV/n region. 
In contrast to antiparticle excesses that can be explained, \eg, by pulsar models or dark matter 
annihilation \citep{Serpico2012,Jie2016Pulsar}, the SOKOL data demand an
enhanced high-energy production of CR deuterons from hadronic interactions. 
We found that no explanation for this measurement
can be provided in terms of standard collisions of CRs with the gas of the ISM. 
As we have shown, a viable solution for this puzzle is the occurrence of nuclear fragmentation inside SNRs,
but this mechanism conflicts with the new \AMS{} data on the \BC{} ratio.  

Thus, if the SOKOL measurement is taken as face value,
we conclude that the sources accelerating helium and protons (thereby producing deuteron) 
may not be the same as those accelerating \C-\N-\O{} nuclei (otherwise producing \Li-\Be-\B{} nuclei), 
and that the former are more efficient in the production and acceleration of secondary particles. 
Under such a scenario, the connection between \BC{} ratio and antiparticle/particle ratios would also be broken,
and thus the \BC{} ratio should not be used to place constraints on the astrophysical antimatter background.
On the other hand, the \dHe{} ratio would still represent a direct diagnostic tool for assessing this background.
\\[0.12cm]
{\footnotesize%
  We are grateful to Tanguy Pierog, Colin Baus, and Ralf Ulrich for the \texttt{CRMC} interface to MC generators.
  J.F. acknowledges support from China Scholarship Council and the Taiwanese Ministry of Science and
  Technology under grants No. 104-2112-M-001-024 and No. 105-2112-M-001-003.
     NT acknowledges support from MAtISSE - \emph{Multichannel Investigation of Solar Modulation Effects in Galactic Cosmic Rays}.
This project has received funding from the European Union's Horizon 2020 research and innovation programme under the Marie Sklodowska-Curie grant agreement No 707543.

}


\end{document}